%% file: paper_guvs_upload.tex
\documentclass[preprint,number,sort&compress]{elsarticle}

\input{packages}

\input{hyperrefpackage}
\input{definitions}

\begin{document}
	
	\begin{frontmatter}
		\title{High-Throughput Mechanical Characterization of Giant Unilamellar Vesicles by Real-Time Deformability Cytometry}
		
		\author[aff1]{Maximilian Kloppe}  
		\ead{Maximilian.Kloppe@math.tu-freiberg.de}
        \author[aff4]{Stefan J. Maurer}
          \author[aff4]{Tobias Abele}
        \author[aff4]{Kerstin Göpfrich}
		\author[aff1,aff2,aff3]{Sebastian Aland \corref{cor1}
		}
		\ead{Sebastian.Aland@math.tu-freiberg.de}
		\cortext[cor1]{Corresponding author}
		\affiliation[aff1]{organization={Institute of Numerical Mathematics and
				Optimization, TU Bergakademie Freiberg},
			addressline={Akademiestr. 6, 09599 Freiberg},
			country={Germany}}
		\affiliation[aff2]{organization={Faculty of Computer Science/Mathematics,
				HTW Dresden},
			addressline={Friedrich-List-Platz 1, 01069 Dresden},
			country={Germany}}
   	\affiliation[aff3]{organization={Center for Systems Biology Dresden},
			addressline={Pfotenhauerstr. 108, 01307 Dresden},
			country={Germany}}
		\affiliation[aff4]{organization={Heidelberg University, Center for Molecular Biology of Heidelberg University (ZMBH)},
			addressline={Berliner Str. 45; D-69120 Heidelberg},
			country={Germany}}
		\begin{abstract}
			Real-time deformability cytometry (RT-DC) enables high-throughput, contact-free mechanical characterization of soft microscopic objects. 
            Here we apply this technique to giant unilamellar vesicles (GUVs).  
            To interpret vesicle deformation in RT-DC, we present a simulation-based model taking into account the area expansion modulus as the dominant mechanical parameter. Using phase-field simulations over a wide parameter space, we find GUV deformation to depend linearly on GUV area. 
            Based on these results, we derive two complementary fitting strategies for extracting the area expansion modulus $K$ from RT-DC data: a direct model-based fit for single-vesicle characterization and a noise-resistant collective approach that enables robust population-level estimates. Furthermore, we introduce a combined fitting method that integrates both approaches to filter outliers and improve accuracy in heterogeneous or noisy datasets. All methods scale across varying flow rates, channel geometries and buffer viscosities, and produce predictions of $K$ consistent with literature values for different lipid compositions. Compared to traditional techniques such as micropipette aspiration, our approach offers orders of magnitude higher throughput without mechanical contact, making it particularly suitable for GUV population studies. Beyond mechanical phenotyping, this framework opens new avenues for sorting vesicle populations based on membrane mechanics, a capability of growing interest in synthetic biology and soft matter research.
		\end{abstract}
		
		
	\end{frontmatter}

\section{Introduction}
Giant unilamellar vesicles (GUVs) serve as versatile model membrane systems in biophysical research for studying biological processes such as lipid phase separation \cite{wesolowska2009giant}, membrane fusion \cite{dimova2016electrodeformation} and protein-lipid interactions \cite{WUBSHET2023550, Sezgin_Schwille_2012}. In addition, GUVs are used in biomedical applications such as targeted drug delivery \cite{ALLEN201336} or synthetic cell design \cite{VanDeCauter_GUVs_for_artificial_cells_2023}.
Their versatility stems from their cell-mimicking size (diameter $>1\, \si{\micro\meter}$), which allows for direct observation of the membrane responses to external factors such as ions, molecules, or hydrodynamic flows under a microscope \cite{Dimova_2019}. In addition, GUVs share many structural similarities with cellular membranes, while providing precise control over factors such as membrane composition, mechanics and functionality \cite{WUBSHET2023550,montes2007giant,Dimova_2019,ali2020generation}.

Mechanical parameters, including deformability, elasticity, and membrane tension are fundamental factors influencing the behavior of GUVs and their various applications. These properties govern how the vesicles undergo dynamic shape changes, maintain structural integrity under external forces, and respond to environmental stresses. At the same time, the mechanical parameters are closely linked to the vesicle composition and are particularly influenced by factors such as lipid types, bilayer thickness and the presence of proteins \cite{RAWICZ20084725, Schaefer_2015}. The mechanical properties of cells, on the other hand, are often attributed to their cytoskeleton rather than the membrane itself. GUVs, in presence and absence of cytoskeletal elements, provide the unique opportunity to decouple these effects \cite{FINK2024}.

However, determining mechanical parameters of GUVs experimentally remains challenging and time consuming. For example, micropipette aspiration experiments \cite{RAWICZ20084725}, response measurements to compression between parallel plates \cite{SCHAEFER_2013} or indentation with a conical AFM-tip \cite{Schaefer_2015} have been used to measure the elastic behaviour of GUVs. A common limitation of these methods is their relatively slow throughput, as they typically allow for the characterization of only one vesicle at a time. 

Microfluidic deformability cytometry offers rapid characterization  of thousands of GUVs within minutes, potentially enabling ultra-fast screening of large vesicle populations. 

One advanced method is real-time deformability cytometry (RT-DC), which provides high-throughput, real-time measurements by subjecting the vesicles to controlled flow forces within a microfluidic channel \cite{OTTO_2015}. As the vesicles traverse the channel, their deformation can be measured and analyzed to potentially quantify the mechanical parameters of GUVs such as elasticity and membrane tension. As an added advantage, the GUVs are not in contact with surfaces during the measurement, which, depending on the type of surface interaction can strongly influence the measurement and make the comparison of different conditions challenging.

Predictions of mechanical properties in RT-DC rely on matching experimentally observed cell deformation with theoretical models. For cells, such theoretical models exist and have been used extensively with RT-DC to study the mechanical properties of cells, allowing rapid phenotyping and mechanical profiling \cite{MIETKE20152023,MMokbel_Numerical_Simulation_of_RTDC,OTTO_2015, rosendahl2018real,Fregin_dynamic_RTDC_2019, Wittwer_LookUpRTDC,WittwerReichelAland_2022}.
  
These models consider elastic and viscoelastic materials and are able to link shape deformation with concrete elastic moduli. 
However, all models consider either volumetric structures (i.e. solid bodies) or elastic shells. Lipid vesicle membranes are neither and hence cannot be evaluated with the current models.

Here, we provide a theoretical method to describe GUV deformation  and to measure lipid elastic properties in an ultrafast manner in RT-DC. 
To this end, we present first numerical simulation results of fluid vesicles in RT-DC channels. 
To eliminate the influence of experimental noise we introduce a new fitting algorithm which allows real-time extraction of the area expansion modulus of a vesicle sample despite errors in the imaging data. 
Finally, we illustrate the generality of the approach to deliver accurate estimates when using RT-DC setups of arbitrary flow rates, channel sizes and buffer viscosities. 

\section{Results}
\subsection{Deformation of a GUV in a microfluidic channel}
Lipid vesicles like GUVs are fluid filled compartments with a bilayer of densely packed lipid molecules at their surface. The lipid molecules (in the liquid disordered state) can freely move in the tangential direction along the surface, which implies a vanishing surface shear modulus. On the other hand, the lipid layer resists bending, areal compression and stretching. It has been shown that the stiffness against bending deformations is negligible in RT-DC measurements due to the large forces exerted \cite{MMokbel_Numerical_Simulation_of_RTDC}. However, the layer's resistance to compression and stretching contributes significantly. The corresponding area expansion modulus $K$ is hence the dominant mechanical parameter and the only determinant of vesicle shapes. 

\begin{figure}
	\centering
	\includegraphics[width=1.0\linewidth]{"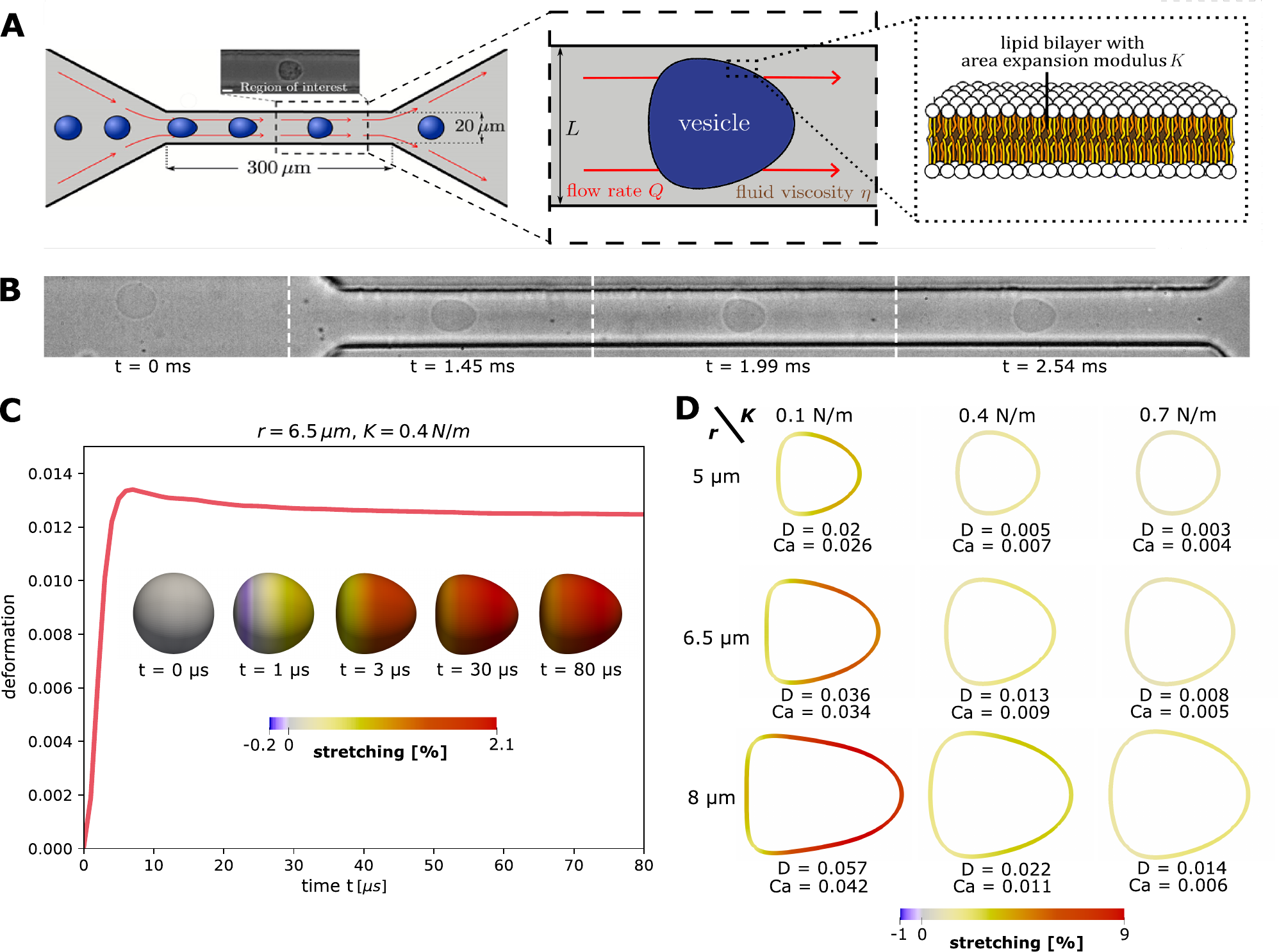"}
	\caption{\textbf{A:} Microfluidic setup of RT-DC and illustration of the physical model. A GUV with an area expansion modulus $K$ is immersed in a surrounding liquid of viscosity $\eta$ and traverses a channel of side length $L$ at a flow rate $Q$. \textbf{B:} Snapshots for RT-DC measurements of GUVs showing a spherical initial state and a bullet-shaped stationary state at the end portion of the channel. \textbf{C:} Simulation of a single GUV. Exemplary evolution for initially spherical vesicle of radius $r=6.5\, \si{\micro \metre}$ and area expansion modulus $K=0.4 \, \si{\newton \per \metre}$, colored by amount of surface stretching (see main text). \textbf{D}: Stationary deformation and approximate capillary numbers for different $r$ and $K$ colored by amount of surface stretching. Simulation parameters: $Q' = 0.04 \, \si{\micro \litre \per \s}, L' = 20 \, \si{\micro \metre}, \eta'= 0.015 \, \si {\Pa \s}$.}
	\label{fig:deformation time plot}
\end{figure}

To extract $K$ from RT-DC image data, we simulate GUV deformation in dependence of $K$. 
We therefore model GUVs as initially spherical thin, freely-shearable elastic surfaces with radius $r$ immersed in a buffer liquid of viscosity $\eta$, within an RT-DC channel with a square-shaped cross-section of side length $L$, see Fig.~\ref{fig:deformation time plot}A. 
A flow rate $Q$ is imposed to set the GUV in motion, such that it subsequently deforms by pressure and shear forces until it assumes a stationary state. 
The full numerical method is based on a phase-field method presented in \cite{KLOPPE2024117090}, with more details to be found in the methods section \ref{sec: numerical method}.
The used simulations parameters ($Q' = 0.04 \, \si{\micro \litre \per \s}, L' = 20 \, \si{\micro \metre}, \eta'= 0.015 \, \si {\Pa \s}$) can be scaled to generalize results to arbitrary $Q, L, \eta$ (see \cref{sec:dependence of E modulus}).

\cref{fig:deformation time plot}C shows an exemplary evolution of GUV deformation.
The initially spherical surface is stretched (and transiently compressed) due to pressure and shear forces until the vesicle reaches a stationary bullet-shape, similar to the shapes observed for RT-DC measurements of GUVs (see \cref{fig:deformation time plot}B) or cells \cite{MMokbel_Numerical_Simulation_of_RTDC}. To quantify deformation we use the measure 
\begin{equation} \label{eq: D defintion}
    D = 1 - {2 \sqrt{A \pi}}/{P},
\end{equation} where $A$ denotes the area and $P$ the perimeter of the two-dimensional image of the GUV shape, as it would be seen by a camera. Note that $D = 0$ for a circle and $D > 0$ for any other shape.
Note that the GUV relaxes back to its spherical shape once it leaves the channel.
Along with the time evolution of $D$, we illustrate the  relative local stretching of the GUV surface, $(\hat{A}-\hat{A}_0)/\hat{A}_0$, with $\hat{A}$ being the area of a surface element in the current configuration and $\hat{A}_0$ the area of this element at the initial time. Hence, stretch values $>0$ and $<0$ correspond to stretching and compression of the surface, respectively, see \cref{fig:deformation time plot}C.

An illustration of the stationary deformation shows that larger and softer GUVs are more stretched, while smaller GUVs with higher $K$ remain close to spherical (\cref{fig:deformation time plot}D). 
The maximum stretching is observed at the lateral front of the vesicle and reaches in our simulations up to 9\% for $r=8\, \si{\micro \metre}$, $K=0.1 \, \si{\newton \per \metre}$. 
This is approximately in the range of stretching tolerated by real lipid bilayers, which is typically limited to a few percents (on the order of $5 \%$ \cite{Ngassam21,Lipowsky_22}) depending on the lipid type. 

For larger stretching, the lipid bilayer may rupture -- a process which is not included in our simulations.

Additionally, we computed for each simulation a rough estimate of the capillary number $Ca = \eta \cdot \dot{\gamma} \cdot r / K$ as a measure of the ratio between the external viscous stress and the membrane's resistance to area expansion, see \cref{fig:deformation time plot}D.
Using the simulation parameters $Q' = 0.04 \, \si{\micro \litre \per \s}, L' = 20 \, \si{\micro \metre}, \eta'= 0.015 \, \si {\Pa \s}$ gives an approximate shear rate of $\dot{\gamma} \approx 35000 \, \si{\per \s}$. 
We find that the computed capillary numbers are fairly close to the deformation values, confirming that GUVs deform in the RT-DC setting primarily in the area-dilatational regime, where the membrane's resistance is dictated mostly by its area expansion modulus $K$.

Note that small fluctuations and excess area are common in GUVs \cite{Aleksanyan31122023}. 
Computing analogously the capillary numbers $Ca_\kappa = \eta \cdot \dot{\gamma} \cdot r^3 / \kappa$ based on the bending rigidity $\kappa$, with $\kappa$ typically being in the order of $10^{-19} \, \si{\newton \metre}$ for GUVs \cite{RAWICZ2000}, gives values $>10^5$, implying that bending contributions are negligible and possibly available excess area is smoothed out immediately. 
This is also confirmed by considering the observed deformation energies, which are $> 10^6 k_B T$, much larger than typical energies of fluctuations.

\subsection{Dependence on elastic modulus} \label{sec:dependence of E modulus}
The goal of this work is to describe the GUV deformation according to its mechanical properties, i.~e. the area expansion elastic modulus $K$, for a given sample of GUVs from RT-DC experiments. The elastic behaviour of lipid bilayers differs strongly depending on the lipid compositions, with values ranging from
$K \in \left[0.15,0.26\right]  \, \si{\newton \per \metre}$ (for DOPC vesicles) \cite{lu2016membrane,RAWICZ2000} 
to $K \in \left[1.56,3.61\right]  \, \si{\newton \per \metre}$ (for $1:1$ SM/Chol vesicles) \cite{McIntosh_1992,RAWICZ20084725}. 

\begin{figure}
	\centering
	\includegraphics[width=1.0\linewidth]{"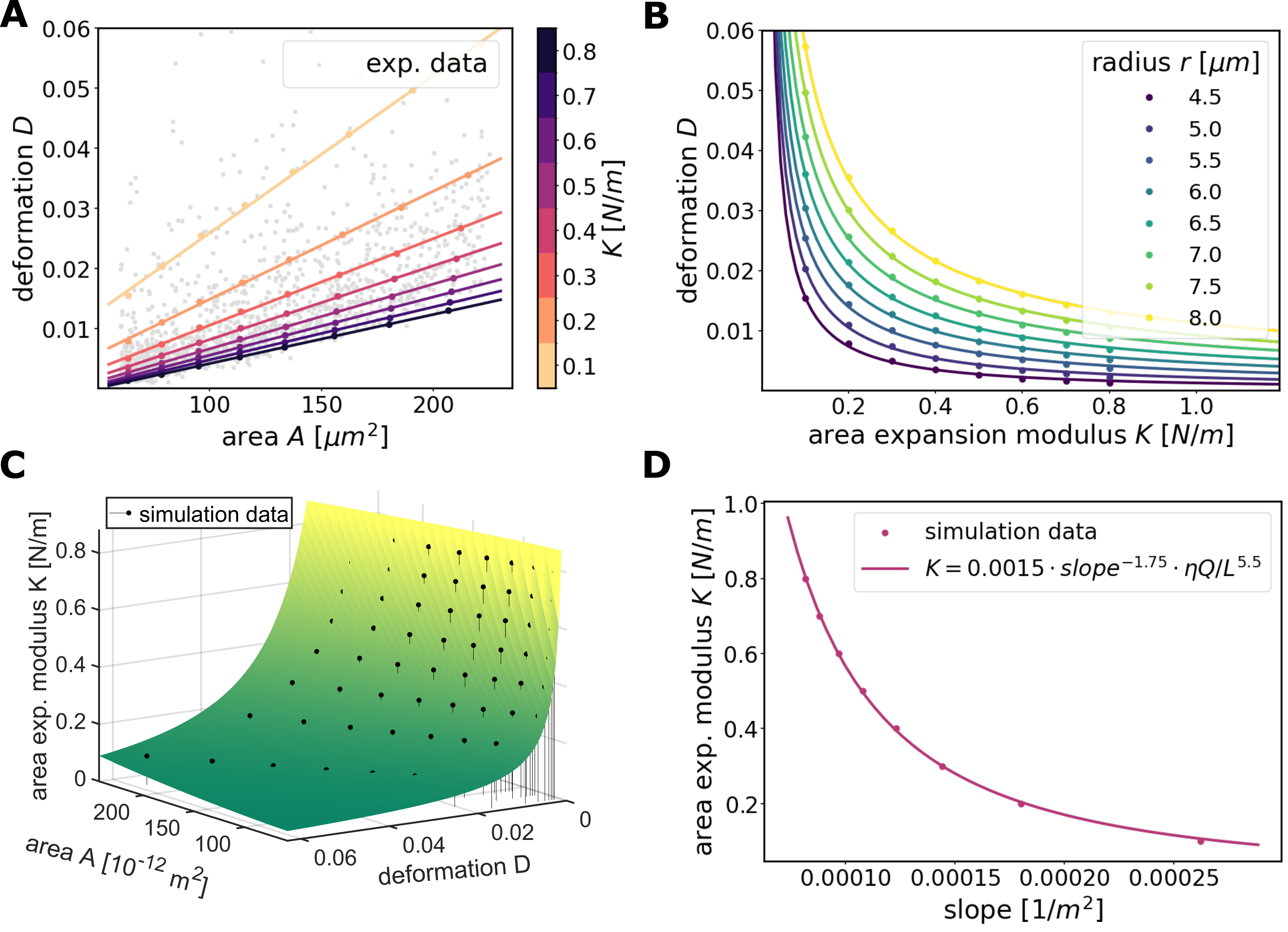"}
	\caption{Simulation data fits. \textbf{A}: Simulation data (dots) fitted by linear functions (lines) illustrating the linear relation between area and deformation. Exemplary experimental data showing an almost linear relationship in GUV measurements as well. \textbf{B}: Simulation data (dots) with exponential fits (lines) illustrating the relation between area expansion modulus and deformation. \textbf{C}: Function fit for the relation between $K$, area and deformation based on \cref{eq:K func final} (direct fitting). \textbf{D}: Fitted function between $K$ and the slope of the linear simulation curves used for the collective fitting method. Parameters: $Q' = 0.04 \, \si{\micro \litre \per \s}, L' = 20 \, \si{\micro \metre}, \eta'= 0.015 \, \si {\Pa \s}$. 
    }
	\label{fig: simulation data}
\end{figure}

Performing simulations for a wide range of elastic moduli and GUV sizes, we find that the deformation in the stationary state depends almost linearly on the GUV size, see \cref{fig: simulation data}A. Linear fits through the simulation data are included to further confirm this observation. Furthermore, an exemplary set of experimentally gathered RT-DC measurements of GUVs is depicted, showing an almost linear relation between $D$ and $A$ for the majority of the data as well.  We note that the almost linear relation differs significantly from RT-DC results which modeled the deformation of volumetric elastic bodies \cite{MMokbel_Numerical_Simulation_of_RTDC} or elastic shells \cite{MMokbel_Numerical_Simulation_of_RTDC,KLOPPE2024117090} instead of fluid vesicle surfaces. 
The dependency between $D$ and $K$ can be well described by a power law $D = \alpha K ^\beta$, \cref{fig: simulation data}B. The good agreement of fits and simulation data is confirmed by the coefficient of determination $R^2 > 0.99$ for all curves.
Using these fitted functions now allows for extrapolations outside of the used parameter range, s.~t. also vesicles with smaller or larger expected values of $K$ can be treated. 

Moreover, the quasi-stationary state of the GUVs is determined by a single non-dimensional parameter $K \cdot (L')^2 / (\eta' \cdot Q')$ \cite{MIETKE20152023}. Hence, analogously to the numerical model presented in \cite{MMokbel_Numerical_Simulation_of_RTDC} our computed results (see \cref{fig: simulation data}A) can be transformed to other flow rates, fluid viscosities and channel sizes. 
More precisely, a change of the parameters from $\eta',Q',L'$ to the values $\eta,Q,L$ can be realized by simply rescaling the area expansion modulus $K$ with factor $\left(\eta Q (L')^2\right)/\left(\eta' Q' L^2\right)$
and rescaling the vesicle area with a factor $(L'/L)^2$. 

\subsubsection{Direct fitting} \label{sec: direct fitting}
Combining both observed dependencies from \cref{fig: simulation data}(A+B), we postulate the following fitting function to describe an explicit relationship between $A$, $D$ and $K$: 

\begin{equation}\label{eq:K fit ansatz}
    \hat{K} = \gamma \cdot D^{-\beta \hat{A}^{\alpha}} \cdot \hat{A}^{-\delta},
\end{equation}
with the dimensionless quantities $\hat{K} = K \cdot L^2/(\eta Q)$ and $\hat{A} = A / L^2$.
The four dimensionless unknowns $\alpha, \beta, \gamma, \delta$ are obtained by a least squares fit with the simulation data (i.e. minimizing the squared difference between $\hat{K}$ from Eq.~\ref{eq:K fit ansatz} and the scaled, dimensionless values from the simulation). We obtain an optimal fit for $\alpha \approx 0.5$, $\beta \approx 2.16$, $\gamma \approx 3.40 \cdot 10^{-4}$, $\delta \approx 1$.
For arbitrary experimental parameters $\eta, Q, L$ which may differ from the ones used in our simulations ($\eta'= 0.015 \, \si {\Pa \s}$, $Q' = 0.04 \, \si{\micro \litre \per \s}$, $L' = 20 \, \si{\micro \metre}$) and when fixing $\alpha = 0.5$ and $\delta = 1$ we obtain the final formula
\begin{equation}\label{eq:K func final}
	K = \dfrac{\gamma \cdot \eta \cdot Q}{A\cdot D^{\beta \sqrt{\frac{A}{{L}^2}}}}.
\end{equation} 

\cref{eq:K func final} now allows to extract the area expansion modulus $K$ directly for each vesicle imaged in RT-DC. The formula is valid for any flow rate, any medium viscosity, and any channel size, as long as the deformation $D$ and the relative area $A/L^2$ are fairly in the range of the here presented simulations ($D\approx\in [0, 0.06], A/L^2\approx \in [0.12, 0.56]$).
\cref{fig: simulation data}C shows that the fitting function describes the data quite well with a coefficient of determination of $R^2 \approx 0.97$ and a sum of squared residuals of $0.091$. 

However, the high-throughput analysis of GUVs in RT-DC is subject to significant data scatter stemming from both inherent physical heterogeneity of the vesicles and various measurement errors. These errors include noise from the imaging system (e.g., low contrast, focus problems, electronic noise) which corrupts the precise determination of the vesicle contour, and additional perturbations (e.g. hydrodynamic interactions of nearby vesicles, flow instabilities, temperature fluctuations) which randomly modulates the effective hydrodynamic stress.
Averaging the obtained $K$-values can cancel out noise. 
To get a representative, statistically less biased value for the whole data set we will introduce a second fitting method in the following section.

\subsubsection{Noise independent collective fitting}\label{sec: collective fitting}

Knowing that the theoretical deformation $D$ depends linearly on the vesicle area $A$ (cf.~\cref{fig: simulation data}A), we find that the slope of the area-deformation curves is a robust marker for the area expansion modulus $K$. 
In fact, postulating the power law $\hat{K} = a \cdot (slope\cdot L^2)^{-b}$ we find an excellent fit ($R^2>0.999$) for the dimensionless parameters $a \approx  0.0015$ and $b \approx 1.75$, see \cref{fig: simulation data}D.
Hence, the formula
\begin{equation}\label{eq: k slope relation}
	K = 0.0015 \cdot (slope\cdot L^2)^{-1.75} \cdot \dfrac{\eta Q}{L^{2}},
\end{equation} 
robustly relates the area-deformation slope of data points to their expansion modulus $K$ for any given $Q, L$, and $\eta$. 

This insight opens the possibility to averaging out measurement errors directly in the data. The precise idea is to apply a linear regression to the raw experimental data in the area-deformation diagram, and then use the obtained slope to extract the area expansion modulus $K$ via Eq.~\eqref{eq: k slope relation}. 
This approach, which we will refer to as collective fitting, promises a more robust identification of $K$, since noise is averaged out directly on the data set. Note, that this simple way of averaging the data is only enabled, by the observed linearity between theoretical area-deformation curves, which does not apply to RT-DC of volumetric elastic bodies. 

We point out, that this idea to first average, then compute $K$, is obviously different from the approach in the last chapter (compute K for each data point, then average over all $K$). In fact, due to the nonlinear dependency of $K$ and $D$ (cf. \cref{eq:K func final}), Jensen's inequality applies, which states that the estimations of both variants must be different. 


With \cref{eq: k slope relation} we can directly infer $K$ from the entire given experimental data set by simply evaluating the slope of a linear fit through the data. Analogously to \cref{eq:K func final}, \cref{eq: k slope relation} is universal and can be used for arbitrary flow rates, buffer viscosities and channel sizes as long as the obtained slopes and relative vesicle areas ($A/L^2$) remain in the range performed in this study.

\subsection{Extraction of dilational elasticity from DOPC vesicles} \label{sec: extraction of K}

Next, we compared our theoretical results to experimental data. RT-DC was carried out by flushing electroformed GUVs of 100\% DOPC through a standard RT-DC chip with a 20 $\mu$m crossection at a flow rate $Q = 0.04 \, \si{\micro\liter \per \second}$ and a corresponding buffer viscosity $\eta \approx 6.3 \, \si{\milli\pascal \second} $, which was estimated using the Buyukurganci-2022 model \cite{buyukurganci22} for a $0.59 \%$ MC-PBS solution (CellCarrier B). 
GUVs were imaged using high-speed brightfield microscopy and their contour was detected. Considering the imaged area and the perimeter of the contour, the deformation $D$ of the GUVs, as a typical output of RT-DC measurements, was calculated using \cref{eq: D defintion} and plotted against the GUV area, \cref{fig:DOPC_chol}A (left).

Due to the short passage times of GUVs ($\sim 1$ms), we observe almost constant GUV volume across the channel (volume change $<1$\% in the first $200 \, \si{\micro \meter}$ behind the entry of the channel), see \cref{sec: RT-DC Experiments}. 

Computing the slope of the linear fit through this data enabled us to compute via Eq.~\eqref{eq: k slope relation} an area expansion modulus of $K=0.186 \, \si{\newton \per \metre}$ for the GUV sample.
Alternatively, using the direct fitting, we computed $K$ for every data point. The corresponding histogram shows a peak around $K=0.19 \, \si{\newton \per \metre}$, \cref{fig:DOPC_chol}A (middle).
Hence, the extracted $K$ values of both the direct and the collective methods are in perfect agreement with the literature values ($K \in [0.15,0.26] \, \si{\newton \per \metre}$, \cite{lu2016membrane,RAWICZ2000}).

Since the measured deformation $D$, is inherently non-negative, the error distribution should be best approximated by a Lognormal distribution. 
Given the power law relationship between $K$ and $D$ (cf. Eq.~\ref{eq:K func final}), the expansion modulus $K$ should also follow a Lognormal distribution.
However, the histogram of $K$, when plotted on a logarithmic scale (or, equivalently, the probability density of $\ln K$), exhibits a clear asymmetry around the peak, deviating from the expected symmetric Gaussian form. 
This asymmetry suggests the presence of two distinct data populations. 
Closer inspection of data points corresponding to very low $K$ (high $D$) revealed those to be unreliable measurements or artifacts (e.g., vesicle doublets or misplaced contour lines). 
We therefore concluded that the dataset contains two statistically separable groups: the true physical measurements (usable GUVs) and the unreliable outliers.
To statistically separate these populations, we model the observed probability density of $\ln K$ using a Gaussian Mixture Model (GMM), assuming $\ln{K} \sim \omega_{A} \, \mathcal{N}(\ln{K_A}, \sigma_{A})+ \omega_{O} \, \mathcal{N}(\ln{K_O}, \sigma_{O})$ with mixture weights $\omega_{A}$ and $\omega_{B}$. 
The resulting GMM provides an excellent fit to the  observed probability density (Figure \ref{fig:DOPC_chol}A middle, orange curve). 
The expansion modulus extracted from the dominant component, $K_A=0.19 \, \si{\newton \per \metre}$, shows robust agreement with the previously reported values.

While both methods (direct and collective fitting) give equally good results for the analyzed data set, the presence of large data scatter and unreliable measurements (outliers) may corrupt the extracted $K$ values for smaller sample sets.
We therefore propose a combined method using the statistically unbiased collective fitting after largely eliminating outliers with the direct fitting approach, as described in the next section.

\subsubsection{Combined fitting gives robust estimates of area expansion modulus}\label{sec: combined fitting}
As described before, the presence of wrongly measured outliers is observed and expected, even after data postprocessing (see \cref{subsec: Deformation deviation correction}). The corresponding variations in deformation might strongly influence the predictions, especially for smaller vesicle populations.

To improve results in such cases, we propose a combined fitting strategy: 
First, remaining outliers are filtered out of the data set. Using the Gaussian Mixture Model we classify data points by calculating how likely the sample belongs to each of the mixture components (posterior probability of each class given the point) and assigning them to the component with the highest probability.  
Hence, the given data set is divided into two clearly distinct clusters. 
The cluster with the lower elasticity (inferred from higher deformations) is marked as outliers and eliminated from the further analysis. 
In a second step, applying the collective fitting method to the remaining data points gives a noise independent prediction of the area expansion modulus of the vesicles, see \cref{fig:DOPC_chol}A (right). 
The obtained modulus $K=0.18 \, \si{\newton \per \metre}$ is again comparable to literature values \cite{lu2016membrane}. 

The approximately equal results for all three fitting strategies (direct, collective, and combined) indicate robustness of all these approaches for the large data set used. 
However, given the inherent variability present for smaller datasets or stiffer GUVs, the extracted $K$-values show moderate differences between these models. 
The combined fitting method is anticipated to be the most reliable, as it effectively filters outliers and exhibits greater insensitivity to noise. 
Therefore, all subsequent data analysis and reported parameters will be based on the combined fitting strategy.

\subsubsection{Extraction for different lipid compositions}\label{sec: extraction of K subsec}
We illustrate the power of collective fitting with supplementary filtering step for different lipid compositions in the following. It is known that the lipid composition influences $K$, however literature values vary. RT-DC would be the first high-throughput method to measure $K$ of vesicle populations. 

First we analyzed different data sets of $100\%$ DOPC vesicles for the two flow rates $Q = 0.04 \, \si{\micro\liter \per \second}$ and $Q = 0.08 \, \si{\micro\liter \per \second}$.
The usage of different flow rates also influences the viscosity of the surrounding fluid due to shear thinning. To account for this, we computed the effective viscosity  again via the model from \cite{buyukurganci22} and obtained $\eta \approx 6.3 \, \si{\milli\pascal \second}$ for $Q = 0.04 \, \si{\micro\liter \per \second}$ and  $\eta \approx 4.7 \, \si{\milli\pascal \second}$ for $Q = 0.08 \, \si{\micro\liter \per \second}$.

\begin{figure}
 	\centering
 	\includegraphics[width = 1.0 \linewidth]{"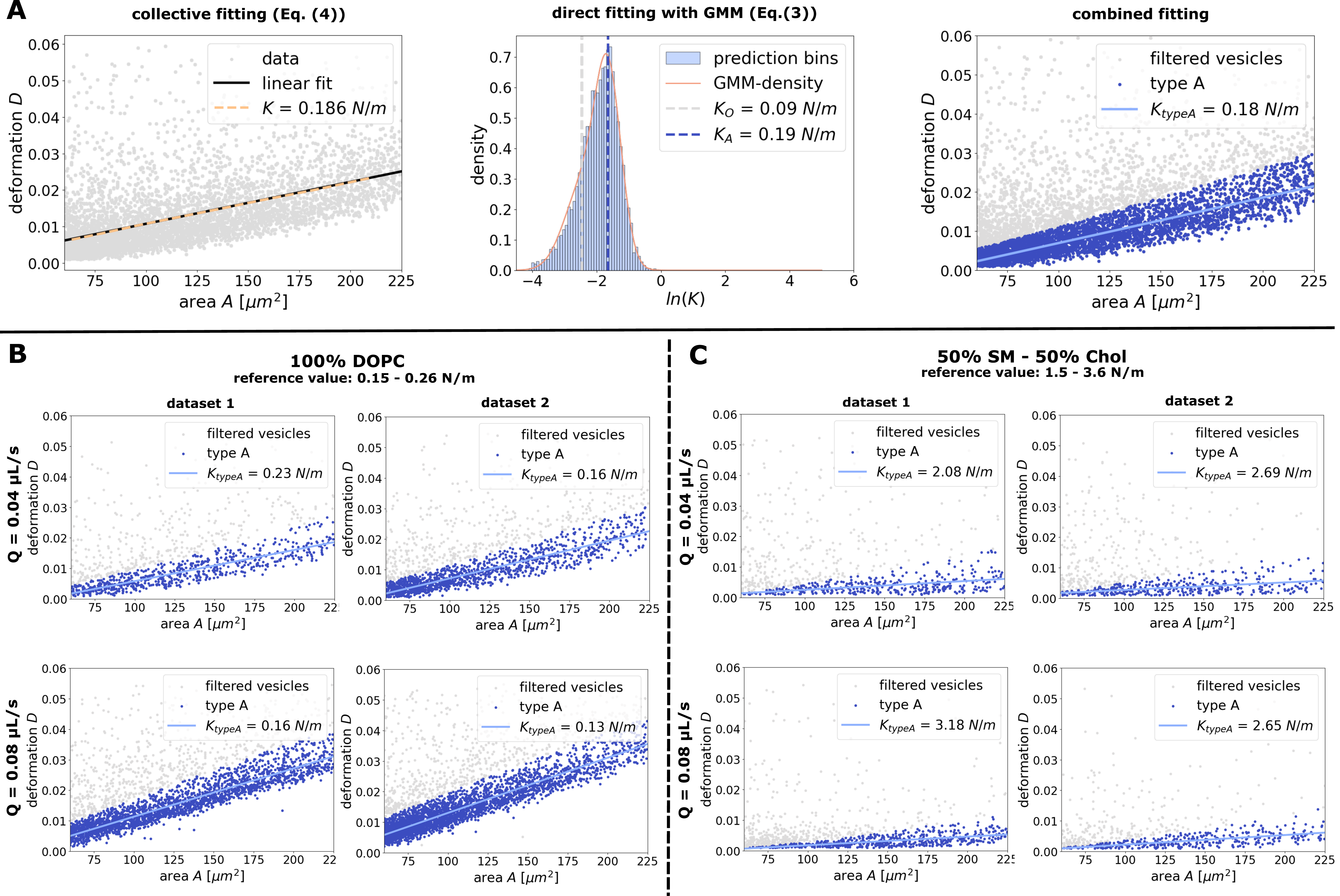"}
 	\caption{Extraction of $K$ using the combined fitting approach. \textbf{A:} Comparison of the three different fitting methods for a given data set with snapshots of measured GUVs (here pure DOPC vesicles). \textbf{Left:}  Extraction of $K$ using collective fitting (\cref{eq: k slope relation}). \textbf{Middle:} Extraction of $K$ using direct fitting (\cref{eq:K func final}). Computation and visualization of Gaussian mixture model (GMM) with two components. \textbf{Right:} Classification of input data based on the Gaussian mixture model (filtering for outliers). Refitting and noise elimination for the usable data with collective fitting approach (\cref{eq: k slope relation}). 
  \textbf{B:} Extraction of $K$ for different sets of pure DOPC vesicles measured at different flow rates ($Q = 0.04 \, \si{\micro\liter \per \second}$, $Q = 0.08 \, \si{\micro\liter \per \second}$).
  \textbf{C:} Extraction of $K$ for different sets of binary composed SM-Chol vesicles ($1:1$) measured at different flow rates ($Q = 0.04 \, \si{\micro\liter \per \second}$, $Q = 0.08 \, \si{\micro\liter \per \second}$).
  }
 	\label{fig:DOPC_chol}
 \end{figure}

Following the combined fitting procedure from \cref{sec: combined fitting}, we computed a linear fit through the filtered data and computed its slope for two independently measured data sets.
Using Eq.~\eqref{eq: k slope relation} we obtained elastic moduli between $K = 0.13 \, \si{\newton \per \metre}$ and $K = 0.23 \, \si{\newton \per \metre}$), see \cref{fig:DOPC_chol}B. 
All the predictions are fairly in the expected range derived from micropipette aspiration ($K \in [0.15,0.26] \, \si{\newton \per \metre}$, \cite{lu2016membrane,RAWICZ2000}).

In a second set of experiments we studied SM-Chol vesicles (1:1 mixture). These vesicles are known to be much stiffer then DOPC vesicles with an expected value for $K$ in the range $K \approx \in [1.5,3.6] \, \si{\newton \per \metre}$ \cite{McIntosh_1992,RAWICZ20084725}. Again we extracted the area expansion modulus for multiple data sets at the two flow rates $Q = 0.04 \, \si{\micro\liter \per \second}$ and $Q = 0.08 \, \si{\micro\liter \per \second}$, see \cref{fig:DOPC_chol}C. 
The extracted values for $K$ are in the range of $K \approx \in [2.08,3.18] \, \si{\newton \per \metre}$, which agrees perfectly with the expected reference range.

Moreover, the variations of extracted $K$-values between the different data sets are for both sets of experiments comparable to those between different flow rates, highlighting the universality of the extraction method to give comparable results across different flow rates and buffer viscosities.

In contrast to the controlled membrane tension environment used in aspiration experiments, GUVs in RT-DC are not initially subjected to a consistent prestress. As a result, small excess area, common in GUVs \cite{Aleksanyan31122023}, must be smoothed out first, leading to an initial deformation response that does not reflect the true elastic stretching of the lipid bilayer. 
Consequently, the apparent stiffness deduced from RT-DC experiments tends to underestimate the intrinsic area expansion modulus $K$ of the membrane.
However, the observation that varying flow rates result in differing deformations ($D$) but yield similar elastic moduli ($K$) provides a strong justification that the influence of membrane fluctuations and excess area is secondary. 
If significant membrane excess area were present, it would stretch out already at very low flow rates, introducing a dominant, flow-rate-independent component to the deformation response. This would in turn lead to calculated modulus $K$ increasing almost linearly with the flow rate (cf. Eqs.~\eqref{eq:K func final}-\eqref{eq: k slope relation}), which we do clearly not observe.

\subsection{Analyzing mixed sets of differently composed vesicles} \label{sec: mixed sets}
The previous examples illustrate that we can extract the area expansion modulus $K$ for a given set of identical GUVs using the collective fitting with a supplementary filtering step.
In the following, we will apply the combined fitting method to classify and identify mixed sets of different GUVs.  

\begin{figure}
	\centering
        \includegraphics[width = 1.0 \linewidth]{"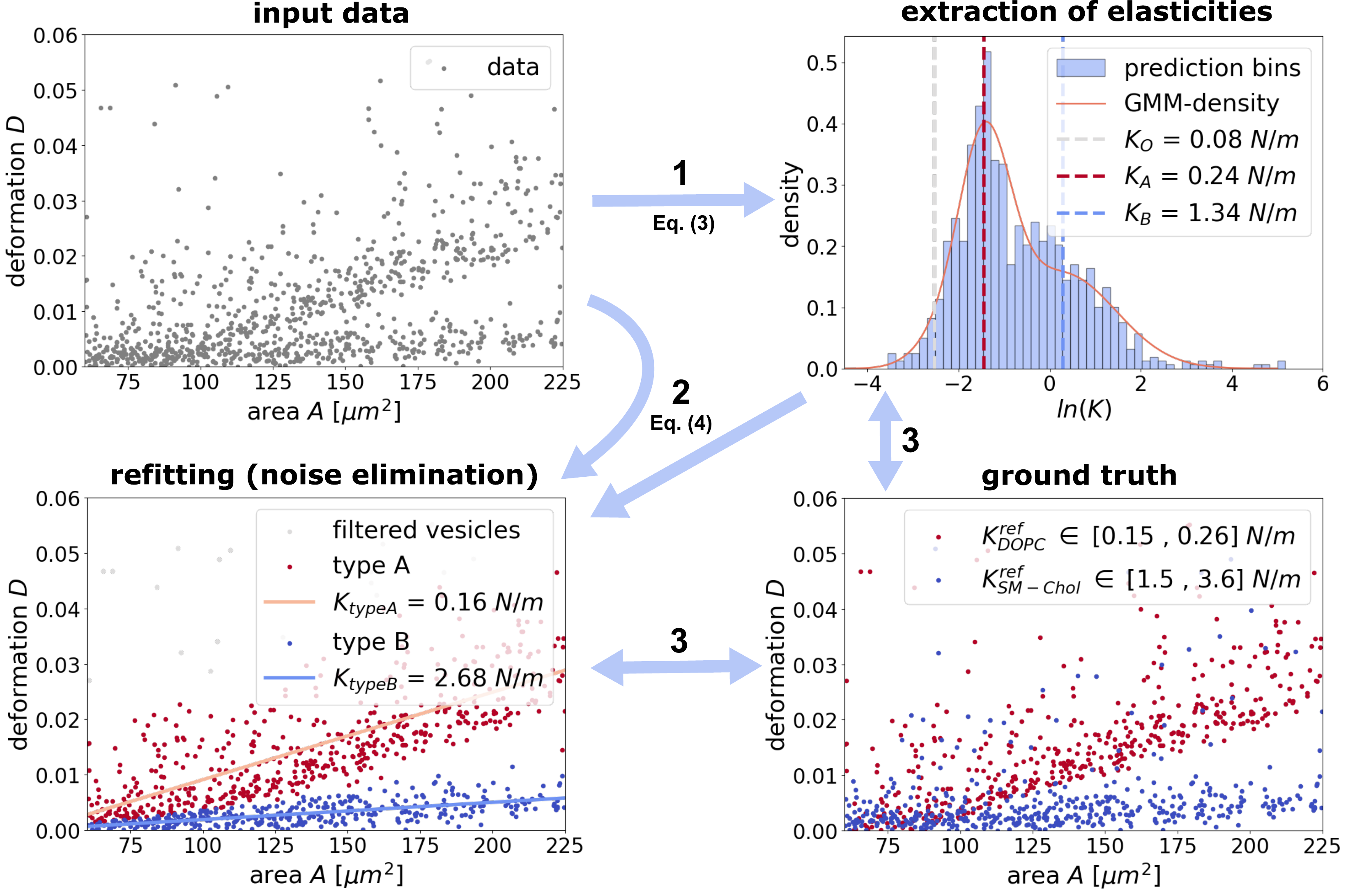"}
	\caption{
    Extraction of area expansion modulus $K$ for a mixture of $1000$ DOPC- and $1000$ SM-Chol-vesicles.
 \textbf{Top (left)}: Given (merged) data set. \textbf{Top (right) - Step 1:} Extraction of $K$ using direct fitting (\cref{eq:K func final}). Computation and visualization of Gaussian mixture model with three components. \textbf{Bottom (left) - Step 2:} Classification of input data based on the Gaussian mixture model. Refitting and noise elimination for the two classes with collective fitting approach (\cref{eq: k slope relation}). \textbf{Bottom (right) - Step 3:} Comparison of data classifications and $K$-extractions with given ground truth (DOPC-vesicles are colored in red, SM-Chol-vesicles are colored in blue), containing possible unreliably results.
 }
	\label{fig:DOPC_SM_mix}
\end{figure}

We consider a data set consisting of $1000$ randomly selected samples from a given DOPC set as well as $1000$ samples from a set of binary composed SM-Chol-vesicles measured at the same flow rate $Q = 0.08 \, \si{\micro\liter \per \second}$ with areas in the range $A \in [60,225] \, \si{\micro\square\metre} $ and deformations $D \le 0.06$. \cref{fig:DOPC_SM_mix} (top left) shows the merged data set. As can be seen from the plot, the data is spread across the whole domain, while forming two distinguishable clusters as well as an amount of outliers. This suggests a mixture of (at least) two different types of GUVs within the data.  

The composition of the data set is assumed to be unknown in this example. Instead we will distinguish the given data by the actual mechanical behaviour. This is again done by applying the direct fitting method using function \eqref{eq:K func final}, giving an estimate for the area expansion modulus for any vesicle in the given data set. 
Visualizing all estimates in a histogram (see \cref{fig:DOPC_SM_mix} (top right)) illustrates that the data (indeed) consists of at least two different types of vesicles. Note that the predictions are again plotted on a logarithmic scale to visualize variations properly. 
Following the ideas from \cref{sec: combined fitting} while assuming three classes (type A, type B, outliers O) within our data, we use a mixture of three Gaussian distributions to classify the distribution of the data, i.~e.
\begin{equation*}
    \ln{K} \sim \omega_{A} \, \mathcal{N}(\ln{K_{A}}, \sigma_{A})+ \omega_{B} \, \mathcal{N}(\ln{K_{B}}, \sigma_{B}) + \omega_{O} \, \mathcal{N}(\ln{K_{O}}, \sigma_{O}).
\end{equation*}

For the computed histogram we get a Gaussian mixture distribution with mean values $K_O \approx 0.08 \, \si{\newton \per \metre}$, $K_A \approx 0.24 \, \si{\newton \per \metre}$ and $K_B  \approx 1.34 \, \si{\newton \per \metre} $ 
and  corresponding standard deviations of $\sigma_{A} \approx 0.58$, $\sigma_{B} \approx 1.17 $  and $\sigma_{O} \approx 0.55 $ respectively. Also the mixture weights are $\omega_{A} \approx 0.50$, $\omega_{B} \approx 0.46$ and $\omega_{O} \approx 0.04$. 
The ratio of the dominant weights ($\omega_{A}$ and $\omega_{B}$) accurately indicates the real half-and-half mixing ratio of the two populations.

In a second step, we use these results to classify the given data to any of the three classes based on the probability distribution and apply again collective fitting to eliminate noise within the data and to get more precise $K$-predictions for the data set.

\cref{fig:DOPC_SM_mix} (bottom left) shows the data set after classification and filtering for outliers. Also we applied collective fitting for both vesicle types. For type A we get an estimate of $K_{typeA} \approx 0.16 \, \si{\newton \per \metre}$.  
For the vesicles of type B we get an estimate of $K_{typeB} \approx 2.68 \, \si{\newton \per \metre}$.

Finally, since the vesicle compositions of the given data are known, we can compare in a third step the estimates of our fitting methods with the ground truth.  \cref{fig:DOPC_SM_mix} (bottom right) shows the merged data set with $1000$ DOPC-vesicles (colored in red) and $1000$ SM-Chol-vesicles (colored in blue) and their respective reference values for $K$ \cite{RAWICZ20084725}.

Firstly, we observe that the direct fitting with function \eqref{eq:K func final} in step one gives clearly distinguishable estimates for $K$. This suggests that function \eqref{eq:K func final}, despite its sensitivity to noise, can be used to indicate the presence of different types of GUVs in a data set and even help to distinguish them by their estimates of the area expansion modulus.

Despite possible misclassification of some of the GUVs due to their similar measured behavior, we get reasonable predictions for both DOPC ($K_{typeA} \approx 0.16 \, \si{\newton \per \metre}$ vs. $K_{ref} \approx \in  [0.15,0.26] \, \si{\newton \per \metre}$) and SM-Chol vesicles ($K_{typeB} \approx 2.68 \, \si{\newton \per \metre}$ vs. $K_{ref} \approx \in [1.5,3.6] \, \si{\newton \per \metre}$) when using the combined fitting.

Note, that the extracted $K$-values from the initial direct fitting method deviate from the final results of the combined fitting ($K_A \approx 0.24 \, \si{\newton \per \metre}$ vs. $K_{typeA} \approx 0.16 \, \si{\newton \per \metre}$ and $K_B  \approx 1.34 \, \si{\newton \per \metre} $ vs. $K_{typeB} \approx 2.68 \, \si{\newton \per \metre}$). 
This discrepancy is a direct consequence of the different statistical treatments: the GMM-derived component means reflect a soft, posterior-weighted expectation that accounts for contributions of all data points according to their classification probabilities (hence points in regions of overlap influence multiple components), whereas the combined method uses a hard assignment (each vesicle is assigned to the most likely class) and then extracts $K$ for each class. 

This example illustrates that the combined fitting approach can also be applied to detect the presence of different types of GUVs within a data set and extract the corresponding elastic moduli.

\section{Conclusion}

In this study, we applied Real-Time Deformability Cytometry to characterize Giant Unilamellar Vesicles. 
To quantify GUV mechanical properties, we introduced a numerical model to describe GUV deformation during this contact-free high-throughput microfluidic process. 
Including the dilational elasticity of the vesicle surface as the dominant mechanical element, we find that vesicles assume a stationary bullet shape, akin to experimentally observed shapes of cells. 
Interestingly, we find that GUV deformation depends linearly on the imaged area of the GUV, with a slope that depends on the area expansion modulus $K$.
Combined with experimental measurements of various vesicle types, we provided evidence that $K$ is the primary factor governing vesicles deformation, while bending contributions are negligible and effects of possibly available excess area are small. 

Based on simulation results for a large number of parameter combinations, we developed fitting functions to extract $K$. 
We showed that our results enable reasonable predictions of $K$ for arbitrary flow rates, buffer viscosities and channel sizes, even compensating noise.  
The direct fitting approach (\cref{sec: direct fitting}) enables $K$-predictions for single vesicles from imaging their deformation and size. 
While this approach is more sensitive to  
noisy image data, it can be used to filter for outliers in wide spread data (see \cref{sec: extraction of K subsec}) or classify different types of vesicles in a mixed data set, as shown in \cref{sec: mixed sets}, as long as the vesicles are distinguishable by their mechanical behavior.

To obtain statistically less biased 
estimates, we introduced the collective fitting method (see \cref{sec: collective fitting}) based on the slope of area-deformation curves. This method yields the averaged area expansion modulus for the whole data set. It works well for noisy image data as it combines information of a whole collective of vesicle deformations to eliminate noise effects in the measured vesicle contours. 

The presence of strong variations within the experimental data still influences the results of the collective fitting method. To eliminate the influence of these outliers and ensure the applicability of our method, we introduced the combined fitting approach (see \cref{sec: combined fitting}). It combines the advantages of both fitting methods leading to more reliable and robust predictions, as shown in \cref{sec: extraction of K subsec}. Also, we showed the applicability of this method for mixed sets of differently composed vesicles (see Sec.~\ref{sec: mixed sets}).

All fitting approaches are presented here such that they scale for arbitrary experimental settings, i.~e. arbitrary flow rates, channel sizes and buffer viscosities. 
Compared to micropipette aspiration, RT-DC tends to slightly underestimate the area expansion modulus, probably due to the absence of a controlled membrane tension environment.
Nevertheless, RT-DC reproduces the mechanical properties of vesicles reasonably well and provides significantly higher throughput in a contact-free manner. Therefore, we believe that it is particularly suited for GUV population analysis. RT-DC provides additional information, like GUV numbers and size distributions which are useful for downstream experiments. For synthetic cell research and beyond, it will become exciting to sort GUVs dependent on their mechanical properties which would lie within the capabilities of the existing RT-DC method \cite{nawaz2020intelligent, nawaz2023image}.

In summary, we successfully demonstrated the application of RT-DC for characterizing GUVs. Beyond precisely measuring the area expansion modulus ($K$), this study establishes the capability to simultaneously quantifying vesicle size, membrane viscosity, and mechanical heterogeneity, thus paving the way for rapid screening and sorting of GUVs by RT-DC.

\section{Methods}
\subsection{Numerical simulations} \label{sec: numerical method}
The presented simulations describe a single initially spherical vesicle with radius $r$ and area expansion modulus $K$ placed in the middle of a square channel of length $L=20\mu$m. 
Approximating the vesicle shapes as  rotationally symmetric and using the concept of the equivalent channel radius \cite{MIETKE20152023}, the simulations are reduced to a 2D axisymmetric setting with a cylindrical channel of radius $R = 10.94 \, \si{\micro \metre}$.
This approximation by axisymmetric shapes was analyzed in the context of RT-DC for volumetric elastic bodies in \cite{Wittwer_LookUpRTDC}. Based on their results we expect accurate results for small vesicles, but a slight underestimation of up to 10\% in the area expansion modulus of the largest vesicles used in this study (area $\approx 200\mu$m). 
The computational domain is prescribed to move with the vesicle barycenter to well resolve the flows and forces around the vesicle at all times. 
In this case, as shown in \cite{MMokbel_Numerical_Simulation_of_RTDC} a channel length of $40 \, \si{\micro \metre}$ is sufficient to provide accurate vesicle deformation. Accordingly, the computational domain is chosen $\Omega = \left[0,40 \right] \, \si{\micro \m} \times \left[0,10.94 \right] \, \si{\micro \m}$. 

The axisymmetric model is based on a phase-field description of elastic surfaces in fluid flow developed in \cite{KLOPPE2024117090}. The resulting system  is discretized using a semi-implicit Euler scheme in time and a  Finite-Element-scheme in space (see \cite[Sec. 4]{KLOPPE2024117090}). The timestep size is chosen as $0.05 \, \si{\micro \s}$. The grid size is approximately $0.5 \, \si{\micro \metre}$ in the fluid phases, while we use an adaptive refinement strategy to ensure a sufficient smooth resolution of the used phase-field function (by at least five grid points at any time around the interface) to describe the vesicle membrane.  

The laminar flow with magnitude $0.21 \, \si{\metre \per \s}$ is induced through appropriate Dirichlet boundary conditions for the velocity field \textbf{v}. This ensures the typical flow rate $Q' = 0.04 \, \si{\micro\litre \per \s}$.  No slip boundary conditions are imposed on the channel wall as well. 
In the simulations the densities and viscosities within the vesicles as well as in the surrounding fluid are set equally as $\rho_0 = \rho_1 = 1000 \, \si{\kilo \g \per  \metre \cubed}$ and $\eta' = \eta_0 = \eta_1 = 0.015 \, \si {\Pa \s}$. 
Note, that stationary results are independent of the viscosity of the vesicles's interior $\eta_1$ since there is (almost) no internal flow in the stationary state. 

The problems are discretized in the finite element toolbox AMDiS \cite{PraetoriusAMDiS,AMDIS_2007,Witkowski_adaptive_parallel_FEM}. The resulting linear system of equations is solved using the direct solver UMFPACK \cite{Davis_UMFPACK}.

\subsection{Fitting procedure}
Throughout this work we used multiple function fits to describe different relationships within the simulation and experimental data. Each fitting problem is formulated as an unconstrained Least-Squares optimization problem and solved based on the Levenberg–Marquardt algorithm (LMA, \cite{More_Levenberg_Marquardt}), which is a standard solver in many software packages. In case of the one-dimensional function fits, e.~g. the relation between $K$ and the slope of the simulation curves, we directly used the computed parameters. For the two-dimensional fit based on \cref{eq:K fit ansatz} we used a two-stepped fitting procedure to reduce the complexity of the problem while maintaining the accuracy of the solution. It turned out, that fixing the parameters $\alpha = 0.5$ and $\delta=1$ gives as good results as solving for four unknowns, which led to the more elegant function \eqref{eq:K func final}.

The effective viscosities necessary for all our calculations were estimated using the experimentally established model from  \cite{buyukurganci22} assuming CellCarrier B (Zellmechanik Dresden) as a buffer medium. 
The slight dilution of CellCarrier B used in our experimental setup (\cref{sec: RT-DC Experiments}) tends to marginally influence fluid viscosity, but has only a negligible effect on the predicted $K$. 
Local variations in shear thinning were shown to have negligible effect on the mechanical characterization by RT-DC at the considered flow rates \cite{Wittwer_LookUpRTDC}.

\subsection{GUV preparation}
Giant unilamellar vesicles (GUVs) composed of 100\% 18:1 (\(\Delta\) 9-Cis) PC (DOPC), or 50\% 18:0 SM (d18:1/18:0)(SM) and 50\% Chol (Avanti Polar Lipids) were prepared via electroformation as described in \cite{Abele}, using a Vesicle Prep Pro (Nanion Technologies GmbH). In brief, 40 $\mu$l of the lipid mixture dissolved at 5 mM in chloroform were spread on the conductive side of an indium tin oxide (ITO) coated glass slide (Visiontek Systems Ltd) and dried under vacuum for 15 min. Using vacuum grease (Carl Roth), a rubber ring was placed on top of the lipid coated slide. The ring was filled with 275 $\mu$l of a 320 mM sucrose solution (Sigma-Aldrich) and a second ITO slide was used to create a closed chamber with the conductive side facing downward. Next, the electrodes of the Vesicle Prep Pro were connected to the conductive side of the ITO slides and the \textit{Standard} program was started, generating an AC field of 3V at 5Hz for 2h. The temperature was set to 70 \textdegree C for GUVs containing SM and to 37 \textdegree C for GUVs containing DOPC.

\subsection{RT-DC measurements} \label{sec: RT-DC Experiments}
For experimental determination of vesicle deformation, an AcCellerator (Zellmechanik Dresden) combined with a high-speed CMOS camera (MC1362, Microtron) on an inverted microscope (AxioObserver, Carl Zeiss AG) with an xy-stage and a 40 ×/0.65 objective was used.  $150 \, \si{\micro \liter}$ of 320mM Glucose solution was added to $90 \, \si{\micro \liter}$ of electroformed vesicles in 320mM Sucrose solution in an Eppendorf tube. After 10 min vesicles were collected from the bottom of the tube and mixed with CellCarrier B (Zellmechanik Dresden) in a 1:9 volumetric ratio. 1 ml glass syringes filled with CellCarrier B or GUVs in CellCarrier B were mounted into neMESYS syringe pumps (Cetoni) and connected to a rectangular 20-$\si{\micro \meter}$-wide channel (Flic20, Zellmechanik Dresden) polydimethylsiloxane chip, via PTFE tubing (S181012, BOLA) and a PTFE plunger (SETonic). The chip was mounted onto the microscope stage and measurements were conducted using Shape-In 2 (Zellmechanik Dresden) at a mean temperature of $23.7^\circ$C, flow rates of 0.04, 0.06, 0.08 or 0.12 $\, \si{\micro \liter \per \second}$ and a sheath flow three times higher than the sample flow.

Vesicles were measured at the end portion of the squared channel in their stationary state with sufficient distance to the channel exit ($>30 \, \si{\micro\meter}$).

\begin{figure}
	\centering
        \includegraphics[width = 0.8 \linewidth]{"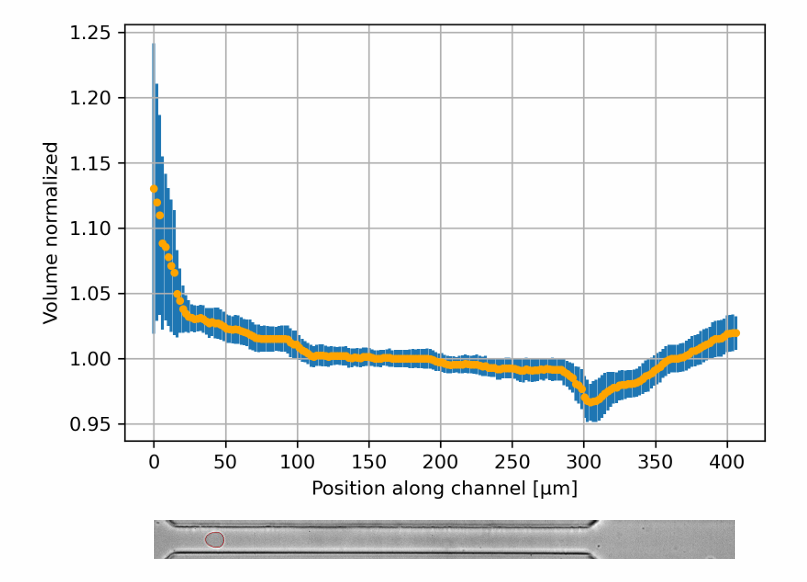"}
	\caption{
    Volume (normalized) over channel position for RT-DC measurements of GUVs (flowing from right to left) showing almost constant GUV volume across the channel.
 }
	\label{fig:volume evolution}
\end{figure}

Moreover, due to the short passage times of GUVs ($\sim 1$ms), we observed almost constant GUV volume across the channel (volume change $<1$\% in the first $200 \, \si{\micro \meter}$ behind the entry of the channel), see \cref{fig:volume evolution}.

\subsection{RT-DC data post processing}\label{subsec: Deformation deviation correction}
It is known that in RT-DC pixelation effects lead to a systematic deformation offset, such that even perfectly circular objects appear slightly deformed when their contours are pictured on a pixel grid. 
Herold  \cite{herold2017mappingdeformationapparentyoungs} quantified this effect for the camera resolution used in our experimental setup ($340 \, \si{\nano \meter} /$pixel) and showed that the deviation decreases with increasing object size but remains non-negligible, particularly for small objects. 
To avoid systematic overestimation of deformation, we apply the published correction function (\cite{herold2017mappingdeformationapparentyoungs}, Eq. 1) to all our data sets prior to further analysis. 
This ensures that our presented results in this work are not strongly biased by pixelation artifacts.
Finally, we restricted the experimental data to the range of our simulation data, i.~e. we just considered data points for $A \in [60,225] \, \si{\micro\square\metre} $ and $D \le 0.1$.

Moreover, the contour detection of giant unilamellar vesicles in RT-DC is susceptible to segmentation errors, often stemming from low image contrast or poorly defined vesicle edges. Such inaccuracies yield irregular or non-convex contours, leading to unreliable measurements of vesicle area ($A$) and deformation ($D$).
\\
To robustly eliminate these artifacts, we employed a filtering strategy based on the ratio between convexified and non-convexified vesicle areas.   
This ratio, also known as RT-DC 'porosity', should ideally be close to unity ($\approx 1.0$). Deviations significantly greater than $1.0$ indicate geometric flaws, such as indentations from overlapping vesicles (doublets) or severely jagged contours.
\\
We chose a robust threshold and excluded all data points where the porosity was greater than $1.03$. This criterion effectively removed the majority of missegmented contours and ambiguous outliers from the dataset. While additional parameters, such as aspect ratio and brightness metrics, were also evaluated as quality indicators, the convexity criterion proved to be the most effective single filter for isolating geometrically compromised events.

\subsection*{Acknowledgements}
SA acknowledges support from DFG (grant numbers 328170591 and 417223351). 
The authors gratefully acknowledge computing time on the high-performance computer at the NHR Center of TU Dresden. This center is jointly supported by the Federal Ministry of Education and Research and the state governments participating in the NHR (www.nhr-verein.de/unsere-partner). Also the authors acknowledge computing time on the compute cluster of the Faculty of Mathematics and Computer Science of Technische Universität Bergakademie Freiberg, operated by the computing center (URZ) and funded by the Deutsche Forschungsgemeinschaft (DFG) under DFG grant number 397252409.
K.G. was funded by the Deutsche Forschungsgemeinschaft (DFG) – SFB-1638/1 – 511488495 - P15 and the ERC starting grant ENSYNC (101076997). K.G. further acknowledges funding from the Deutsche Forschungsgemeinschaft (DFG, German Research Foundation) under Germany's Excellence Strategy via the Excellence Cluster 3D Matter Made to Order (EXC-2082/1 -- 390761711).
\bibliographystyle{acm}
\bibliography{literature}

\end{document}

%% file: packages.tex
\usepackage[utf8]{inputenc}
\usepackage[main=english]{babel}
\usepackage[automark]{scrlayer-scrpage}
\usepackage{a4wide}
\usepackage[leqno]{amsmath}
\usepackage{amssymb}
\usepackage{amsfonts}
\usepackage{mathtools}
\usepackage{breqn}
\usepackage{theorem}
\usepackage{array}
\usepackage{booktabs}
\usepackage{colortbl}
\usepackage{dcolumn}
\usepackage{alltt}
\usepackage{paralist}
\usepackage{siunitx}
\usepackage{supertabular}
\DeclareGraphicsExtensions{.pdf,.jpg,.eps}
\usepackage{tabularray}
\usepackage{pifont}
\usepackage{tikz}
\usetikzlibrary{arrows.meta}
\usetikzlibrary{decorations.pathreplacing}
\usepackage[ruled, vlined, algosection]{algorithm2e}
\usepackage[noend]{algpseudocode}
\usepackage{listings}
\usepackage{enumitem}

\usepackage{comment}

\usepackage{caption}
\usepackage{subcaption}

\usepackage{ dsfont }


%% file: hyperrefpackage.tex
\usepackage{hyperref}
\hypersetup{
  linkcolor=magenta,
  filecolor=blue,
  colorlinks=true
}

\usepackage[capitalise]{cleveref}

\crefname{section}{Sec.}{Secs.}
\Crefname{Section}{Sec.}{Secs.}

%% file: definitions.tex
\SetKwInput{KwData}{Gegeben}

\definecolor{darkgreen}{rgb}{0,0.6,0}
\definecolor{gray}{rgb}{0.5,0.5,0.5}
\definecolor{mauve}{rgb}{0.58,0,0.82}



%% file: literature.bib
@article{lu2016membrane,
  title={Membrane mechanical properties of synthetic asymmetric phospholipid vesicles},
  author={Lu, Li and Doak, William J and Schertzer, Jeffrey W and Chiarot, Paul R},
  journal={Soft Matter},
  volume={12},
  number={36},
  pages={7521--7528},
  year={2016},
  publisher={Royal Society of Chemistry}
}

@article{MIETKE20152023,
title = {Extracting Cell Stiffness from Real-Time Deformability Cytometry: Theory and Experiment},
journal = {Biophysical Journal},
volume = {109},
number = {10},
pages = {2023-2036},
year = {2015},
issn = {0006-3495},
doi = {https://doi.org/10.1016/j.bpj.2015.09.006},
url = {https://www.sciencedirect.com/science/article/pii/S000634951500939X},
author = {Alexander Mietke and Oliver Otto and Salvatore Girardo and Philipp Rosendahl and Anna Taubenberger and Stefan Golfier and Elke Ulbricht and Sebastian Aland and Jochen Guck and Elisabeth Fischer-Friedrich}
}

@article{SCHAEFER_2013,
author = {Schäfer, Edith and Kliesch, Torben-Tobias and Janshoff, Andreas},
title = {Mechanical Properties of Giant Liposomes Compressed between Two Parallel Plates: Impact of Artificial Actin Shells},
journal = {Langmuir},
volume = {29},
number = {33},
pages = {10463-10474},
year = {2013},
doi = {10.1021/la401969t},
note ={PMID: 23869855},
URL = { 
        https://doi.org/10.1021/la401969t
},
eprint = { 
        https://doi.org/10.1021/la401969t
}
}

@Article{Schaefer_2015,
author ="Schäfer, Edith and Vache, Marian and Kliesch, Torben-Tobias and Janshoff, Andreas",
title  ="Mechanical response of adherent giant liposomes to indentation with a conical AFM-tip",
journal  ="Soft Matter",
year  ="2015",
volume  ="11",
issue  ="22",
pages  ="4487-4495",
publisher  ="The Royal Society of Chemistry",
doi  ="10.1039/C5SM00191A",
url  ="http://dx.doi.org/10.1039/C5SM00191A"
}

@article{RAWICZ20084725,
title = {Elasticity, Strength, and Water Permeability of Bilayers that Contain Raft Microdomain-Forming Lipids},
journal = {Biophysical Journal},
volume = {94},
number = {12},
pages = {4725-4736},
year = {2008},
issn = {0006-3495},
doi = {https://doi.org/10.1529/biophysj.107.121731},
url = {https://www.sciencedirect.com/science/article/pii/S0006349508703400},
author = {W. Rawicz and B.A. Smith and T.J. McIntosh and S.A. Simon and E. Evans}
}

@article{OTTO_2015,
author = {Otto, Oliver and Rosendahl, Philipp and Mietke, Alexander and Golfier, Stefan and Herold, Christoph and Klaue, Daniel and Girardo, Salvatore and Pagliara, Stefano and Ekpenyong, Andrew and Jacobi, Angela and Wobus, Manja and Töpfner, Nicole and Keyser, Ulrich F. and Mansfeld, Jörg and Fischer-Friedrich, Elisabeth and Guck, Jochen},
title = { Real-time deformability cytometry: on-the-fly cell mechanical phenotyping},
journal = {Nature Methods},
volume = {12},
pages = {199-202},
year = {2015},
doi = {10.1038/nmeth.3281},
URL = { 
        https://doi.org/10.1038/nmeth.3281
}
}

@article{AMDIS_2007,
author = {Vey, Simon and Voigt, Axel},
title = {{AMDiS}: adaptive multidimensional simulations},
journal = {Computing and Visualization in Science},
volume = {10},
pages = {57-67},
year = {2007},
doi = {10.1007/s00791-006-0048-3},
URL = { 
        https://doi.org/10.1007/s00791-006-0048-3
}
}

@misc{PraetoriusAMDiS,
	author = {Praetorius, S.},
	title = {The Adaptive Multi-Dimensional Simulation Toolbox ({AMDiS}), a discretization module on top of the {Dune}
	framework},
	howpublished = {\url{https://gitlab.com/amdis/amdis}}
}

@article{Witkowski_adaptive_parallel_FEM,
	title = {Software concepts and numerical algorithms for a scalable adaptive parallel finite element method},
	journal = {Advances in Computational Mathematics},
	volume = {41},
	pages = {1145--1177},
	year = {2015},
	issn = {1572-9044},
	doi = {https://doi.org/10.1007/s10444-015-9405-4},
	author = {Witkowski, T. and Ling, S. and Praetorius, S. and Voigt, A.}
}

@article{Davis_UMFPACK,
	author = {Davis, Timothy A.},
	title = {Algorithm 832: UMFPACK V4.3---an Unsymmetric-Pattern Multifrontal Method},
	year = {2004},
	issue_date = {June 2004},
	publisher = {Association for Computing Machinery},
	address = {New York, NY, USA},
	volume = {30},
	number = {2},
	issn = {0098-3500},
	url = {https://doi.org/10.1145/992200.992206},
	doi = {10.1145/992200.992206},
	journal = {ACM Trans. Math. Softw.},
	pages = {196–199}
}

@article{KLOPPE2024117090,
title = {A phase-field model of elastic and viscoelastic surfaces in fluids},
journal = {Computer Methods in Applied Mechanics and Engineering},
volume = {428},
pages = {117090},
year = {2024},
issn = {0045-7825},
doi = {https://doi.org/10.1016/j.cma.2024.117090},
url = {https://www.sciencedirect.com/science/article/pii/S0045782524003463},
author = {Maximilian Kloppe and Sebastian Aland}
}

@article{MMokbel_Numerical_Simulation_of_RTDC,
author = {Mokbel, M. and Mokbel, D. and Mietke, A. and Träber, N. and Girardo, S. and Otto, O. and Guck, J. and Aland, S.},
title = {Numerical Simulation of Real-Time Deformability Cytometry To Extract Cell Mechanical Properties},
journal = {ACS Biomaterials Science \& Engineering},
volume = {3},
number = {11},
pages = {2962-2973},
year = {2017},
doi = {10.1021/acsbiomaterials.6b00558},
    note ={PMID: 33418716}
}

@InProceedings{More_Levenberg_Marquardt,
	author="Mor{\'e}, Jorge J.",
	editor="Watson, G. A.",
	title="The Levenberg-Marquardt algorithm: Implementation and theory",
	booktitle="Numerical Analysis",
	year="1978",
	publisher="Springer Berlin Heidelberg",
	address="Berlin, Heidelberg",
	pages="105--116",
	isbn="978-3-540-35972-2"
}

@article{Dimova_2019,
   author = "Dimova, Rumiana",
   title = "Giant Vesicles and Their Use in Assays for Assessing Membrane Phase State, Curvature, Mechanics, and Electrical Properties", 
   journal= "Annual Review of Biophysics",
   year = "2019",
   volume = "48",
   number = "Volume 48, 2019",
   pages = "93-119",
   doi = "https://doi.org/10.1146/annurev-biophys-052118-115342",
   url = "https://www.annualreviews.org/content/journals/10.1146/annurev-biophys-052118-115342",
   publisher = "Annual Reviews",
   issn = "1936-1238",
   type = "Journal Article",
  }

@article{Wittwer_LookUpRTDC,
author ="Wittwer, Lucas Daniel and Reichel, Felix and Müller, Paul and Guck, Jochen and Aland, Sebastian",
title  ="A new hyperelastic lookup table for {RT-DC}",
journal  ="Soft Matter",
year  ="2023",
volume  ="19",
issue  ="11",
pages  ="2064--2073",
publisher  ="The Royal Society of Chemistry",
doi  ="10.1039/D2SM01418A",
url  ="http://dx.doi.org/10.1039/D2SM01418A"
}

@incollection{WittwerReichelAland_2022,
title = {Chapter 3 - Numerical simulation of deformability cytometry: Transport of a biological cell through a microfluidic channel},
editor = {Sid Becker and Andrey V. Kuznetsov and Filippo {de Monte} and Giuseppe Pontrelli and Dan Zhao},
booktitle = {Modeling of Mass Transport Processes in Biological Media},
publisher = {Academic Press},
pages = {33-56},
year = {2022},
isbn = {978-0-443-15765-3},
doi = {https://doi.org/10.1016/B978-0-323-85740-6.00010-8},
url = {https://www.sciencedirect.com/science/article/pii/B9780323857406000108},
author = {Lucas Daniel Wittwer and Felix Reichel and Sebastian Aland}
}

@article{Fregin_dynamic_RTDC_2019,
author ="Fregin, Bob and Czerwinski, Fabian and Biedenweg, Doreen and Girardo, Salvatore and Groß, Stefan and Aurich, Konstanze and Otto, Oliver",
title  ="Dynamic Real-Time Deformability Cytometry - Time-Resolved Mechanical Single Cell Analysis at 100 Cells/s",
journal  ="Biophysical Journal",
year  ="2020",
volume  ="118",
issue  ="3",
pages  ="605a",
publisher  ="Elsevier",
doi  ="10.1016/j.bpj.2019.11.3267",
url  ="https://doi.org/10.1016/j.bpj.2019.11.3267"
}

@article{WUBSHET2023550,
title = {Methods to mechanically perturb and characterize GUV-based minimal cell models},
journal = {Computational and Structural Biotechnology Journal},
volume = {21},
pages = {550-562},
year = {2023},
issn = {2001-0370},
doi = {https://doi.org/10.1016/j.csbj.2022.12.025},
url = {https://www.sciencedirect.com/science/article/pii/S2001037022005839},
author = {Nadab H. Wubshet and Allen P. Liu}
}

@article{VanDeCauter_GUVs_for_artificial_cells_2023,
author = {Van de Cauter, Lori and van Buren, Lennard and Koenderink, Gijsje H. and Ganzinger, Kristina A.},
title = {Exploring Giant Unilamellar Vesicle Production for Artificial Cells — Current Challenges and Future Directions},
journal = {Small Methods},
volume = {7},
number = {12},
pages = {2300416},
doi = {https://doi.org/10.1002/smtd.202300416},
url = {https://onlinelibrary.wiley.com/doi/abs/10.1002/smtd.202300416},
eprint = {https://onlinelibrary.wiley.com/doi/pdf/10.1002/smtd.202300416},
year = {2023}
}

@article{ALLEN201336,
title = {Liposomal drug delivery systems: From concept to clinical applications},
journal = {Advanced Drug Delivery Reviews},
volume = {65},
number = {1},
pages = {36-48},
year = {2013},
note = {Advanced Drug Delivery: Perspectives and Prospects},
issn = {0169-409X},
doi = {https://doi.org/10.1016/j.addr.2012.09.037},
url = {https://www.sciencedirect.com/science/article/pii/S0169409X12002980},
author = {Theresa M. Allen and Pieter R. Cullis}
}

@article{rosendahl2018real,
  title={Real-time fluorescence and deformability cytometry},
  author={Rosendahl, Philipp and Plak, Katarzyna and Jacobi, Angela and Kraeter, Martin and Toepfner, Nicole and Otto, Oliver and Herold, Christoph and Winzi, Maria and Herbig, Maik and Ge, Yan and others},
  journal={Nature methods},
  volume={15},
  number={5},
  pages={355--358},
  year={2018},
  publisher={Nature Publishing Group US New York}
}

@article{wesolowska2009giant,
  title={Giant unilamellar vesicles-a perfect tool to visualize phase separation and lipid rafts in model systems},
  author={Weso{\l}owska, Olga and Michalak, Krystyna and Maniewska, Jadwiga and Hendrich, Andrzej},
  journal={Acta Biochimica Polonica},
  volume={56},
  number={1},
  pages={33--39},
  year={2009},
  publisher={Polskie Towarzystwo Biochemiczne}
}

@article{Sezgin_Schwille_2012,
author = {Erdinc Sezgin and Petra Schwille},
title = { Model membrane platforms to study protein-membrane interactions },
journal = {Molecular Membrane Biology},
volume = {29},
number = {5},
pages = {144--154},
year = {2012},
publisher = {Taylor \& Francis},
doi = {10.3109/09687688.2012.700490},
URL = {    
        https://doi.org/10.3109/09687688.2012.700490
},
eprint = {     
        https://doi.org/10.3109/09687688.2012.700490
}
}

@article{dimova2016electrodeformation,
  title={Electrodeformation, electroporation, and electrofusion of giant unilamellar vesicles},
  author={Dimova, Rumiana and Riske, Karin A and Damijan, M},
  journal={Handbook of electroporation},
  pages={235--252},
  year={2016},
  publisher={Springer Cham, Switzerland}
}

@article{montes2007giant,
  title={Giant unilamellar vesicles electroformed from native membranes and organic lipid mixtures under physiological conditions},
  author={Montes, L-Ruth and Alonso, Alicia and Goni, Felix M and Bagatolli, Luis A},
  journal={Biophysical journal},
  volume={93},
  number={10},
  pages={3548--3554},
  year={2007},
  publisher={Elsevier}
}

@article{ali2020generation,
  title={Generation of interconnected vesicles in a liposomal cell model},
  author={Ali Doosti, Baharan and Fj{\"a}llborg, Daniel and Kustanovich, Kiryl and Jesorka, Aldo and Cans, Ann-Sofie and Lobovkina, Tatsiana},
  journal={Scientific Reports},
  volume={10},
  number={1},
  pages={14040},
  year={2020},
  publisher={Nature Publishing Group UK London}
}

@article{Ngassam21,
  title={Recurrent dynamics of rupture transitions of giant lipid vesicles at solid surfaces},
  author={Ngassam, Viviane N. and Su, Wan-Chih and Gettel, Douglas L. and Deng, Yawen and Yang, Zexu and Wang-Tomic, Neven and Sharma, Varun P. and Purushothaman, Sowmya and Parikh, Atul N.},
  journal={Biophysical Journal},
  volume={120},
  number={4},
  pages={586--597},
  year={2021},
  publisher={Elsevier},
  doi = {10.1016/j.bpj.2021.01.006},
URL = {    
        https://doi.org/10.1016/j.bpj.2021.01.006
}
}

@article{Lipowsky_22,
author = {Lipowsky, Reinhard},
title = {Remodeling of Membrane Shape and Topology by Curvature Elasticity and Membrane Tension},
journal = {Advanced Biology},
volume = {6},
number = {1},
pages = {2101020},
doi = {https://doi.org/10.1002/adbi.202101020},
url = {https://onlinelibrary.wiley.com/doi/abs/10.1002/adbi.202101020},
eprint = {https://onlinelibrary.wiley.com/doi/pdf/10.1002/adbi.202101020},
year = {2022}
}

@article{FINK2024,
title = {Membrane localization of actin filaments stabilizes giant unilamellar vesicles against external deforming forces},
journal = {European Journal of Cell Biology},
volume = {103},
number = {2},
pages = {151428},
year = {2024},
issn = {0171-9335},
doi = {https://doi.org/10.1016/j.ejcb.2024.151428},
url = {https://www.sciencedirect.com/science/article/pii/S0171933524000451},
author = {Andreas Fink and Sunnatullo Fazliev and Tobias Abele and Joachim P. Spatz and Kerstin Göpfrich and Elisabetta Ada Cavalcanti-Adam}
}

@article{Abele,
author = {Abele, Tobias and Messer, Tobias and Jahnke, Kevin and Hippler, Marc and Bastmeyer, Martin and Wegener, Martin and Göpfrich, Kerstin},
title = {Two-Photon 3D Laser Printing Inside Synthetic Cells},
journal = {Advanced Materials},
volume = {34},
number = {6},
pages = {2106709},
doi = {https://doi.org/10.1002/adma.202106709},
url = {https://advanced.onlinelibrary.wiley.com/doi/abs/10.1002/adma.202106709},
eprint = {https://advanced.onlinelibrary.wiley.com/doi/pdf/10.1002/adma.202106709},
year = {2022}
}

@article{Aleksanyan31122023,
author = {Mina Aleksanyan and Hammad A. Faizi and Maria-Anna Kirmpaki and Petia M. Vlahovska and Karin A. Riske and Rumiana Dimova and},
title = {Assessing membrane material properties from the response of giant unilamellar vesicles to electric fields},
journal = {Advances in Physics: X},
volume = {8},
number = {1},
pages = {2125342},
year = {2023},
publisher = {Taylor \& Francis},
doi = {10.1080/23746149.2022.2125342},
note ={PMID: 36211231},
URL = {    
        https://doi.org/10.1080/23746149.2022.2125342
},
eprint = {     
        https://doi.org/10.1080/23746149.2022.2125342
}

}

@Article{buyukurganci22,
author ="Büyükurgancı, Beyza and Basu, Santanu Kumar and Neuner, Markus and Guck, Jochen and Wierschem, Andreas and Reichel, Felix",
title  ="Shear rheology of methyl cellulose based solutions for cell mechanical measurements at high shear rates",
journal  ="Soft Matter",
year  ="2023",
volume  ="19",
issue  ="9",
pages  ="1739-1748",
publisher  ="The Royal Society of Chemistry",
doi  ="10.1039/D2SM01515C",
url  ="http://dx.doi.org/10.1039/D2SM01515C"
}

@misc{herold2017mappingdeformationapparentyoungs,
      title={Mapping of Deformation to Apparent Young's Modulus in Real-Time Deformability Cytometry}, 
      author={Christoph Herold},
      year={2017},
      eprint={1704.00572},
      archivePrefix={arXiv},
      primaryClass={cond-mat.soft},
      url={https://arxiv.org/abs/1704.00572}, 
}

@article{RAWICZ2000,
title = {Effect of Chain Length and Unsaturation on Elasticity of Lipid Bilayers},
journal = {Biophysical Journal},
volume = {79},
number = {1},
pages = {328-339},
year = {2000},
issn = {0006-3495},
doi = {https://doi.org/10.1016/S0006-3495(00)76295-3},
url = {https://www.sciencedirect.com/science/article/pii/S0006349500762953},
author = {W. Rawicz and K.C. Olbrich and T. McIntosh and D. Needham and E. Evans}
}

@article{McIntosh_1992,
author = {McIntosh, Thomas J. and Simon, Sidney A. and Needham, David and Huang, Ching Hsien},
title = {Structure and cohesive properties of sphingomyelin/cholesterol bilayers},
journal = {Biochemistry},
volume = {31},
number = {7},
pages = {2012-2020},
year = {1992},
doi = {10.1021/bi00122a017},
note ={PMID: 1536844},
URL = {    
        https://doi.org/10.1021/bi00122a017
},
eprint = {     
        https://doi.org/10.1021/bi00122a017
}

}

@article{nawaz2020intelligent,
  title={Intelligent image-based deformation-assisted cell sorting with molecular specificity},
  author={Nawaz, Ahmad Ahsan and Urbanska, Marta and Herbig, Maik and N{\"o}tzel, Martin and Kr{\"a}ter, Martin and Rosendahl, Philipp and Herold, Christoph and Toepfner, Nicole and Kub{\'a}nkov{\'a}, Mark{\'e}ta and Goswami, Ruchi and others},
  journal={Nature Methods},
  volume={17},
  number={6},
  pages={595--599},
  year={2020},
  publisher={Nature Publishing Group US New York}
}

@article{nawaz2023image,
  title={Image-based cell sorting using focused travelling surface acoustic waves},
  author={Nawaz, Ahmad Ahsan and Soteriou, Despina and Xu, Catherine K and Goswami, Ruchi and Herbig, Maik and Guck, Jochen and Girardo, Salvatore},
  journal={Lab on a Chip},
  volume={23},
  number={2},
  pages={372--387},
  year={2023},
  publisher={Royal Society of Chemistry}
}
